\documentclass[letterpaper, 10 pt, conference]{ieeeconf}  % Comment this line out
\IEEEoverridecommandlockouts                              % This command is only
\usepackage{graphicx} % for pdf, bitmapped graphics files
\usepackage{amsmath} % assumes amsmath package installed
\usepackage{amssymb}  % assumes amsmath package installed
\usepackage[dvipsnames]{xcolor}
\usepackage{booktabs} % required horizontal lines in bmatrix environment
\usepackage{caption}
\usepackage{subcaption}

\usepackage[hang,flushmargin]{footmisc} % To remove indentation in footnotes

\usepackage{tikz}
\usetikzlibrary{decorations.pathreplacing,positioning, arrows.meta}
\title{\LARGE \bf
Data-enabled predictive control with instrumental variables: the direct equivalence with subspace predictive control}

\author{Jan-Willem van Wingerden$^{1}$, Sebastiaan P. Mulders$^{1}$, Rogier Dinkla$^{1}$, Tom Oomen$^{1}$ and Michel Verhaegen$^{1}$ % <-this % stops a space
\thanks{This work is part of the research programme: Robust closed-loop wake steering for large densely
space wind farms" with project number 17512, which is financed by the Dutch Research
Council (NWO).).
$^{1}$Delft University of Technology, Delft Center for Systems and Control, Mekelweg 2, 2628 CD Delft, The Netherlands.
        {\tt\small \{J.W.vanWingerden, S.P.Mulders, R.T.O.Dinkla, T.A.E.Oomen, M.Verhaegen\}@tudelft.nl}.}}

\begin{document}

\maketitle
\thispagestyle{empty}
\pagestyle{empty}

%%%%%%%%%%%%%%%%%%%%%%%%%%%%%%%%%%%%%%%%%%%%%%%%%%%%%%%%%%%%%%%%%%%%%%%%%%%%%%%%
\begin{abstract}
Direct data-driven control has attracted substantial interest since it enables optimization-based control without the need for a parametric model. This paper presents a new Instrumental Variable~(IV) approach to Data-enabled Predictive Control (DeePC) that results in favorable noise mitigation properties, and demonstrates the direct equivalence between DeePC and Subspace Predictive Control (SPC). The methodology relies on the derivation of the characteristic equation in DeePC along the lines of subspace identification algorithms. A particular choice of IVs is presented that is uncorrelated with future noise, but at the same time highly correlated with the data matrix. A simulation study demonstrates the improved performance of the proposed algorithm in the presence of process and measurement noise.
% Instead of projecting away the effect of the input we use ...
\end{abstract}
\section{Introduction}
\noindent In an era where data is abundantly available and the scientific fields of Machine Learning (ML) and Artificial Intelligence (AI) progress rapidly, the field of data-driven control has consequentially attracted a significant interest~\cite{de2019formulas}\cite{8960476}. In particular in the field of Model Predictive Control (MPC), Data-enabled Predictive Control (DeePC) has been substantially developed. This algorithm has its origin in Willems' fundamental Lemma~\cite{willems2005note}, which states that all future input-output trajectories of a linear system are parameterized by a sufficiently excited past input-output trajectory~\cite{van2020willems}.

Data-driven control methods can be categorized into \textit{direct} and \textit{indirect} approaches. DeePC is a direct data-driven control framework in the sense that it uses past input-output data to make future decisions without an explicit parametric model~\cite{dorfler2021bridging}. In sharp contrast, indirect data-driven control methods perform a separate, and often computational and time expensive system realisation step. That is, in subspace identification a singular value decomposition is performed to find the model order followed by a linear regression problem~\cite{verhaegen2007filtering}\cite{van2013closed}. The DeePC algorithm combines the system realization and control steps in a single optimization using solely data-matrices, making it a potentially efficient and promising method for the control of unknown systems. %While in indirect data-driven control is a computational expensive system realisation step (\emph{e.g.} in subspace identification you have to perform an SVD to detect the rank followed by a linear regression problem~\cite{verhaegen1994identification}~\cite{van2013closed}) is performed on the data-matrices in direct data-driven control these data-matrices are directly part of the controller synthesis problem.

In the original DeePC paper~\cite{coulson2019data} the algorithm is derived for deterministic LTI systems without taking noise explicitly into account. Later, in~\cite{dorfler2021bridging} and~\cite{hewing2020learning} the algorithm in~\cite{coulson2019data} is extended to address disturbing noise through a regularization term. In practical application, averaging is applied over the past input-output trajectory to mitigate the effect of noise~\cite{jo2022data}. In~\cite{favoreel1999spc} a direct data-driven control algorithm called Subspace Predictive Control (SPC) is developed, that does take noise into account in its problem formulation without the need for regularization.
%This SPC approach is further developed in~\cite{dong2008closed}, where a closed-loop SPC algorithm is presented with higher data efficiency and explicitly takes the causality constraint into account.
%In the field of closed-loop SPC also extensions already exist for  LPV~\cite{dong2009closed} model structures, and for repetitive control~\cite{navalkar2014subspace}.%subspace predictive control (SPC). is a direct data-driven control algorithm that takes noise directly into account in the problem formulation, without the direct need for regularization~\cite{favoreel1999spc}

Several recent papers connect the different algorithms in the field of data-driven control. Research in~\cite{fiedler2021relationship} shows that SPC and DeePC are equivalent for the deterministic case, whereas the work of~\cite{dorfler2021bridging} makes a bridge between different data-driven control techniques. In addition, in~\cite{dorfler2021bridging} additional regularization terms are suggested to exploit the underlying causality and low-rank properties for the SPC algorithm. If measurement and process noise are added, the original DeePC  algorithm depends on additional regularization and/or data-averaging to mitigate the effect of noise~\cite{jo2022data}. The open challenge is to optimally mitigate the effect of noise in such data-driven control algorithms.

Although substantial developments have been made in data-driven control algorithms, at present the effects of noise are not completely addressed. The aim of this paper is to present a new algorithm based on the DeePC framework, now including instrumental variables (IVs), which is a well-established technique to mitigate the effect of noise~\cite{soderstrom2002instrumental}. In addition, full equivalence between SPC and DeePC with IVs is established. To this end, the characteristic equation used in DeePC is first derived from the data-equation used in subspace identification algorithms including measurement and process noise. Then, IVs are introduced to mitigate the negative effect of noise. By formulating the predictive control problem, it becomes evident that the two seemingly different algorithms are equivalent.

The main contributions of this paper are threefold:
\begin{itemize}
    \item The traditional DeePC algorithm is formulated directly in terms of the data-equations that are encountered in subspace identification.
    \item Instrumental variables are introduced and shown to result in favourable noise mitigation properties.
    \item Direct equivalence of IV-based DeePC and SPC is established.
\end{itemize}

This paper is organized as follows. Section~\ref{Sec_modelstructure} introduces the model structure and notation used. In Section~\ref{Section_char_equation}, the characteristic equation of the DeePC algorithm is derived and extended with instrumental variables~(IVs). These IVs reveal a direct equivalence with SPC. In Section~\ref{Section_Predictive_Control}, the data-enabled predictive control problem is set up, followed by simulation results in Section~\ref{Sec_Simulations}. Finally, conclusions are drawn in Section~\ref{Sect_Conclusions}.

%%%%%%%%%%%%%%%%%%%%%%%%%%%%%%%%%%%%%%%%%%%%%%%%%%%%%%%%%%%%%%%%%%%%%%%%%%%%%%%%%%%%
\section{Model structure and notation \label{Sec_modelstructure}}
\noindent This section presents the model structure and the notation used for the derivation of the different algorithms considered in this paper.

\subsection{Model structure}
\noindent The assumed model structure, commonly used in the field of subspace identification, is given by the following Linear Time-Invariant (LTI) discrete-time system in innovation form~\cite{ljung1998system}:
\begin{equation}
\begin{aligned}
x_{k+1}&=&Ax_k&+Bu_k+Ke_k,\\
y_{k}&=&Cx_k&+Du_k+e_k, \label{eqmod}
\end{aligned}
\end{equation}
\noindent where $x_k \in \mathbb{R}^n$, $u_k
\in \mathbb{R}^r$, $e_k \in \mathbb{R}^\ell $, $y_k
\in \mathbb{R}^\ell$, are the state, input, noise, and output vectors, respectively, and $k \in \mathbb{Z}_{\geq 0}$ denotes the discrete time index.
The matrices $A \in \mathbb{R}^{n\times n}$, $B \in \mathbb{R}^{n\times r}$, $K \in
\mathbb{R}^{n\times \ell}$, $C \in \mathbb{R}^{\ell \times n}$, $D \in \mathbb{R}^{\ell \times r}$, are
the respective system, input, Kalman, output, and direct feedthrough matrices. By manipulation of~(\ref{eqmod}), the so-called predictor form is obtained:
\begin{equation}
\begin{aligned}
x_{k+1}&=&\tilde{A}x_k&+\tilde{B}u_k+Ky_k,\\
{y}_{k}&=&Cx_k&+Du_k+e_k, \label{eqmod_innovation}
\end{aligned}
\end{equation}
with $\tilde{A}={A}-{K}C$ and $\tilde{B}={B}-{K}D$. The prediction problem inherently present in a data-enabled predictive control setting can now be formulated as:\vspace{0.2cm}

%\noindent \textbf{Problem statement:} Given a past input trajectory of, $u_k$, and output trajectory, $y_k$, for
%$k=\{i,\ldots, N-1+i\}$, find an unbiased linear predictor that can predict the future outputs $y_k$ as function of $u_k$ for $k=\{N+i,\,\ldots\,, N+f-1+i\}$ where $f \in \mathbb{R}^{+}$ defines the prediction window and $N \in \mathbb{R}^{+}$ the number of samples.

\noindent \textbf{Problem statement:} Given $N$ data instances of an unknown LTI system (\ref{eqmod}), determine a data-driven prediction of the system trajectory as function of the associated control action over a prediction horizon $f$.

\subsection{Assumptions and notation}
\noindent This section introduces the notation and assumptions used throughout this paper, see also~\cite{van2013closed}. First, the block-Hankel matrix is defined as:
\begin{equation}
Y_{i,s,\bar{N}}=\begin{bmatrix}y_i & y_{i+1} & \hdots & y_{i+\bar{N}-1}\\
y_{i+1} & y_{i+2} & \hdots & y_{i+\bar{N}}\\
\vdots & \vdots & \ddots & \vdots\\
y_{i+s-1} & y_{i+s} & \hdots & y_{i+\bar{N}+s-2}\\
\end{bmatrix},
\end{equation}
with $Y_{i,s,\bar{N}} \in \mathbb{R}^{\ell s \times \bar{N}}$, $i \in \mathbb{Z}$,  and $\left\{s,\,\bar{N}\right\}\in \mathbb{Z}_{> 0}$. The block-Hankel matrices $U_{i,s,{\bar{N}}} \in \mathbb{R}^{r s \times \bar{N}}$ and $E_{i,s,\bar{N}} \in \mathbb{R}^{\ell s \times \bar{N}}$ are defined in a similar fashion.  Note that $\bar{N}+s-1$ data samples are used to construct this block-Hankel matrix.

\noindent A block-Toeplitz matrix is defined as follows:
\begin{equation}
H_{(B,D)}=\begin{bmatrix}D & 0 & 0& \hdots & 0\\
CB & D & 0& \hdots & 0\\
\vdots & \vdots & \vdots & \vdots & \vdots \\
CA^{f-2}B & CA^{f-3}B & \hdots & CB &D\\
\end{bmatrix},
\end{equation}
with $H_{(B,D)} \in \mathbb{R}^{\ell f \times rf}$, and likewise $H_{(K,I)}  \in \mathbb{R}^{\ell f \times \ell f}$ with $f \in \mathbb{Z}_{> 0}$.

\noindent Next, the extended controllability matrix is defined:
\begin{equation}
\mathcal{{K}}_{(\tilde{B})}=\begin{bmatrix}\tilde{A}^{p-1} \tilde{B}  & \tilde{A}^{p-2}\tilde{B}  & \hdots& \tilde{B}
\end{bmatrix},
\end{equation}
with $\mathcal{{K}}_{(\tilde{B})}\in \mathbb{R}^{n  \times rp}$, and similarly  $\mathcal{{K}}_{(K)}\in \mathbb{R}^{n  \times \ell p}$ with $p \in \mathbb{Z}_{> 0}$. For presentation reasons we also introduce ${\mathcal{{K}}=\begin{bmatrix}
\mathcal{{K}}_{(\tilde{B})} & \mathcal{{K}}_{(K)}
\end{bmatrix}\in \mathbb{R}^{n  \times (r+\ell)p}}$. The extended observability matrix is defined as:
%\begin{equation}
%\Gamma=\begin{bmatrix}C \\ CA \\ CA^2 \\ \vdots \\ CA^{f-1}
%\end{bmatrix},
%\end{equation}
\begin{equation}
\Gamma=\begin{bmatrix}C \\ CA \\ CA^2 \\ \vdots \\ CA^{f-1}
\end{bmatrix},
\end{equation}
with ${\Gamma}\in \mathbb{R}^{\ell f  \times n}$. Finally, the variable $O$ indicates a zero matrix, $o$ a zero vector, and $I$ the identity matrix, all of appropriate dimensions.
\begin{figure*}
\begin{center}
\includegraphics[width = 1.00\textwidth]{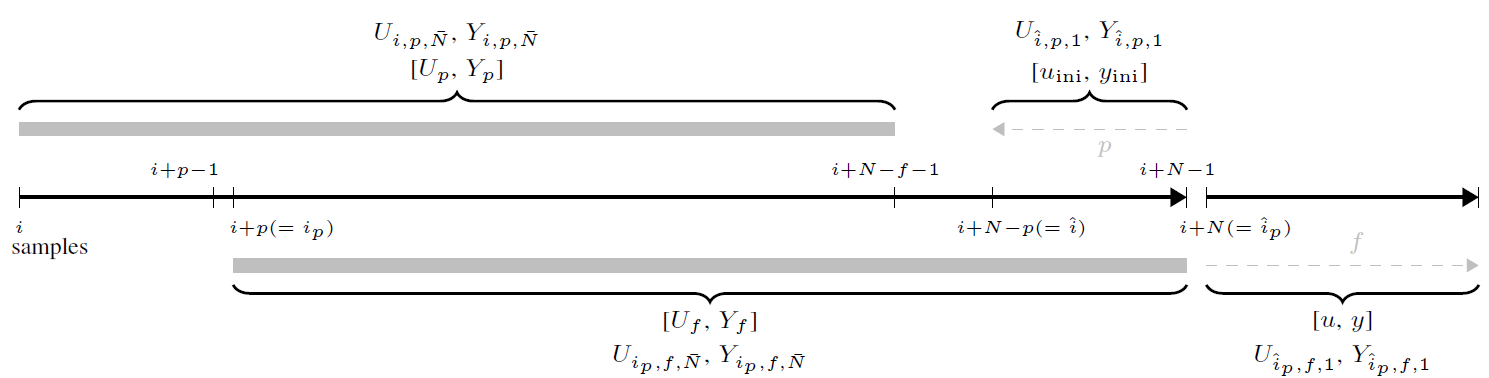}
\end{center}
\vspace*{-0.4cm}
\caption{Visualisation how the data is used within the data-enabled predictive control algorithm. The notation used in this paper is provided, whereas the variables within square brackets correspond to the the notation used in~\cite{coulson2019data}.}
\label{Fig_windows}
\end{figure*}

%%%%%%%%%%%%%%%%%%%%%%%%%%%%%%%%%%%%%%%%%%%%%%%%%%%%%%%%%
\section{The characteristic equation \label{Section_char_equation}}
\noindent In this section, the characteristic equation used in data-enabled predictive control is derived in an alternative way. Section~\ref{sect:data_eq} introduces the data equation. Then, the characteristic equation for DeePC is derived in Section~\ref{sect:Char_DeePC}. Section~\ref{sect:Char_DeePCIV} introduces the concept of instrumental variables and applies them to the characteristic equation. Finally, Section~\ref{sect:SPC} exhibits the resulting equivalence of DeePC and SPC by the application of IVs.
\subsection{The data equation \label{sect:data_eq}}
\noindent An underlying key result that is exploited in both DeePC and subspace identification methods is that the state of a system can be expressed in terms of past input-output data. Forward propagation of the state equation in (\ref{eqmod_innovation}) over $p$ samples leads to:
\begin{eqnarray}
x_{k+p}=\tilde{A}^px_k+\mathcal{K}\begin{bmatrix}U_{k,p,1}\\Y_{k,p,1}\end{bmatrix}.
\end{eqnarray}
With the assumption that $\tilde{A}$ is stable there exist a finite $p$ such that  $||\tilde{A}^p||_F\approx 0$, which is an approximation commonly used in many subspace algorithms~\cite{chiuso2007role}.\footnote{Please note that in the noiseless case there exists a deadbeat observer, $K$, such that $||\tilde{A}^n||_F=0$~\cite{houtzager2009varmax}.} For large enough $p$, the following well-known data equation can be constructed using the definitions from the previous subsection~\cite{van2013closed}:\medmuskip=0mu
\begin{eqnarray}
Y_{i_p,f,\bar{N}}=\Gamma\mathcal{K}
\begin{bmatrix}U_{i,p,\bar{N}}\\Y_{i,p,\bar{N}}\end{bmatrix}+H_{(B,D)}U_{i_p,f,\bar{N}}+H_{(K,I)}E_{i_p,f,\bar{N}},
\label{eq_Data_equation_1}
\end{eqnarray}
with $\bar{N}=N-p-f+1$ and $i_p=i+p$. Using the input-output trajectories from time instance $i$ until $i+N-1$.   This data equation is used in many subspace identification algorithms (\emph{e.g.} N4SID~\cite{van1994n4sid}, PO-MOESP~\cite{verhaegen2007filtering} or PBSID~\cite{van2013closed}). In the PO-MOESP algorithm an orthogonal projection is used to remove the effect of $U_{i_p,f,\bar{N}}$ to isolate the low rank matrix $\Gamma\mathcal{K}$. In the PO-MOESP instrumental variables are used to suppress the effect of the noise. The key idea of the DeePC~\cite{coulson2019data} algorithm is to retain $U_{i_p,f,\bar{N}}$ in the data equation, while the effect of noise was not included in the derivation of the original algorithm.

The data equation~\eqref{eq_Data_equation_1}, connects available data from the so-called past (block-Hankel matrices with a $p$ in the subscript) to the available data in the so-called future (block-Hankel matrices with an $f$ in the subscript). These overlapping data-sets are also illustrated in Fig.~\ref{Fig_windows} with the gray bars. In the DeePC algorithm a second similar data equation is defined which connects the last $p$ input-output samples available in the data set to $f$ unknown input-output samples in the future. This is illustrated in~Fig.~\ref{Fig_windows} by the dashed gray arrows. This second data-equation is given by:
\begin{eqnarray} \medmuskip=0mu
\color{black}Y_{\hat{i}_{p},f,1}\color{black}=\Gamma\mathcal{K}
\begin{bmatrix}U_{\hat{i},p,1}\\Y_{\hat{i},p,1}\end{bmatrix}+H_{(B,D)}\color{black}U_{\hat{i}_{p},f,1}\color{black}+H_{(K,I)}E_{\hat{i}_{p},f,1},
\label{eq_Data_equation_2}
\end{eqnarray}
with $\hat{i}=i+N-p$ and $\hat{i}_p=i+N$. Note that with the current indices  $\color{black}\color{black}Y_{\hat{i}_p,f,1}\color{black}\color{black}$, $\color{black}\color{black}U_{\hat{i}_p,f,1}\color{black}\color{black}$ are unknown vectors and the variables that will be used in the predictive control problem.
%In Fig.~\ref{Fig_windows} the data-matrices used in this paper are graphically presented and related with the notation used in~\cite{coulson2019data}.

%%%%%%%%%%%%%%%%%%%%%%%%%%%%%%%%%%%%%%%%%%%%%
\subsection{The DeePC algorithm characteristic equation \label{sect:Char_DeePC}}
\noindent In this section, the data equation, which is the result of the previous subsection, is used to derive the characteristic equation for DeePC. To this end, first the deterministic LTI case  is considered, whereas the following subsection incorporates noise in the data equation while also introducing~IVs.

Equations~\eqref{eq_Data_equation_1}-\eqref{eq_Data_equation_2} are respectively rewritten for the noise-free case by omitting the noise term as:
\begin{eqnarray}
\begin{bmatrix}\Gamma\mathcal{K}
 & H_{(B,D)} & -I \end{bmatrix}
\begin{bmatrix}\begin{bmatrix}U_{i,p,\bar{N}}\\Y_{i,p,\bar{N}}\end{bmatrix}\\U_{i_p,f,\bar{N}}\\Y_{i_p,f,\bar{N}}\end{bmatrix}&=&O\label{eq_Data_equation_1_DeePC},\\
\begin{bmatrix}\Gamma\mathcal{K}  & H_{(B,D)} & -I \end{bmatrix}
\begin{bmatrix}\begin{bmatrix}U_{\hat{i},p,1}\\Y_{\hat{i},p,1}\end{bmatrix}\\\color{black}\color{black}U_{\hat{i}_p,f,1}\color{black}\\\color{black}Y_{\hat{i}_p,f,1}\color{black}\color{black}\end{bmatrix}&=&o.\label{eq_Data_equation_2_DeePC}
\end{eqnarray}
To obtain the characteristic equation of the DeePC algorithm,~\eqref{eq_Data_equation_1_DeePC} is at the right hand side multiplied with a vector ${\color{black}g\color{black} \in \mathbb{R}^{\bar{N} \times 1}}$. This basically means that linear combinations of input-output trajectories in the available data set are taken. By subtracting \eqref{eq_Data_equation_2_DeePC} from \eqref{eq_Data_equation_1_DeePC} and introducing $g$, we obtain:
\medmuskip=0mu
\begin{eqnarray}
\begin{bmatrix}\Gamma\mathcal{K}  & H_{(B,D)} & -I \end{bmatrix}
\left(\begin{bmatrix}\begin{bmatrix}U_{i,p,\bar{N}}\\Y_{i,p,\bar{N}}\end{bmatrix}\\U_{i_p,f,\bar{N}}\\Y_{i_p,f,\bar{N}}\end{bmatrix}\color{black}g\color{black}-\begin{bmatrix}\begin{bmatrix}U_{\hat{i},p,1}\\Y_{\hat{i},p,1}\end{bmatrix}\\\color{black}\color{black}\color{black}U_{\hat{i}_p,f,1}\color{black}\\\color{black}Y_{\hat{i}_p,f,1}\color{black}\color{black}\color{black}\end{bmatrix}\right)=o.\label{eq_Data_equation_1_DeePC2}
\end{eqnarray}
This result shows that an unknown input-output trajectory can be embedded as a linear combination of available input-output trajectories.
Because the pre-multiplication matrix is full-rank, the following is obtained:
\begin{eqnarray}
\begin{bmatrix}\begin{bmatrix}U_{i,p,\bar{N}}\\Y_{i,p,\bar{N}}\end{bmatrix}\\U_{i_p,f,\bar{N}}\\Y_{i_p,f,\bar{N}}\end{bmatrix}\color{black}g\color{black}=\begin{bmatrix}\begin{bmatrix}U_{\hat{i},p,1}\\Y_{\hat{i},p,1}\end{bmatrix}\\\color{black}\color{black}\color{black}U_{\hat{i}_p,f,1}\color{black}\\\color{black}Y_{\hat{i}_p,f,1}\color{black}\color{black}\color{black}\end{bmatrix},\label{eq_Data_equation_1_DeePC3}
\end{eqnarray}
which can be used in a data-enabled prediction control problem with $\color{black} g$, $\color{black}U_{\hat{i}_p,f,1}\color{black}$ and $\color{black}Y_{\hat{i}_p,f,1}\color{black}$ as decision variables. Although the derivation is different, the result is in accordance with the results presented in~\cite{coulson2019data}.

%Note that in the noiseless case, the conditions of sufficient excitation and a rank of at least $(f+p)r+n$ should hold for the data matrix, \textit{i.e.}, $\bar{N}>(f+p)r+n$. The first part of this condition assumes that the input is persistently excited and thus $Y_{i,p,\bar{N}}$ has rank $pr$ and $U_{i_p,f,\bar{N}}$ rank $fr$. Since there is no noise, a Kalman gain $K$ exists such that $\tilde{A}^n=O$ and thus the matrix $Y_{i,p,\bar{N}}$ has rank $n$.

%Since the last block row of the characteristic equation is already spanned by the other rows.

As already shown in~\cite{coulson2021distributionally}, \eqref{eq_Data_equation_1_DeePC3} can be split into two equations:
\begin{eqnarray}
\begin{bmatrix}\begin{bmatrix}U_{i,p,\bar{N}}\\Y_{i,p,\bar{N}}\end{bmatrix}\\U_{i_p,f,\bar{N}}\end{bmatrix}\color{black}g\color{black}&=&\begin{bmatrix}\begin{bmatrix}U_{\hat{i},p,1}\\Y_{\hat{i},p,1}\end{bmatrix}\\\color{black}U_{\hat{i}_p,f,1}\color{black}\\\end{bmatrix},\label{eq_Data_equation_1_DeePC4}\\
Y_{i_p,f,\bar{N}} \color{black} g \color{black} &=&\color{black}Y_{\hat{i}_p,f,1}\color{black}.
\end{eqnarray}
The data matrix in the left-hand side of (\ref{eq_Data_equation_1_DeePC4}), has dimensions $\left(r+\ell\right)p+rf\,\times\,\bar{N}$. Assuming that the conditions on persistency of excitation are fulfilled, it should hold in the deterministic case that $\bar{N}\geq(r+\ell)p+rf$ to parameterize all possible future trajectories.
%This expression is also valid when there is noise but the rank of the matrix will be $(f+p)r+p\ell$.

In the case of process and measurement noise, the trajectories in the available data set are corrupted by noise and consequently the DeePC algorithm will select linear combinations of these trajectories leading to a reduced performance.
%In the noiseless situation, all future input and output trajectories are parameterized if $\bar{N}= (r+\ell)p+rf$. If $\bar{N}>(r+\ell)p+rf$, linear combinations of the already existing trajectories are added to the data set. However, in the case that new noise corrupted trajectories are added. Of course, by specifying $\bar{N}=(r+\ell)p+rf$, a unique result is obtained, but the effect of noise is not reduced.
As proposed in~\cite{coulson2019data}, noise can be mitigated through regularization, or as proposed in~\cite{jo2022data}, mitigated by columnwise data-averaging such that the augmented block-Hankel matrix is square ($\bar{N}=(r+\ell)p+rf$). The central idea in the present paper is to employ IVs for a systematic mitigation of noise, as is presented in the next section.
%In the case of noise, the future trajectories will be linear combination of noise corrupted trajectories leading to a loss in performance. The approaches of averaging the trajectories and/or regularization can be beneficial in the presence of noise~\cite{coulson2019data}.

%Another approach that is new in the field of data-enabled predictive control is the use of instrumental variables, and obviate the need for the former two mentioned methods. The derivation of the DeePC algorithm with instrumental variables to reduce the effect of noise is outlined in the next section.
%%%%%%%%%%%%%%%%%%%%%%%%%%%%%%%%%%%%%%%%%%%%%%%%%%%%%%%%%%%%%%%%%%%%%%%%%
\subsection{The DeePC algorithm with Instrumental Variables\label{sect:Char_DeePCIV}}
\noindent In this section, the main result of DeePC with instrumental variables that systematically mitigate noise is presented. Equation~(\ref{eq_Data_equation_1}) is again taken as a starting point for the derivation including noise and is rewritten as:
\begin{eqnarray}
\begin{bmatrix}\Gamma\mathcal{K} & H_{(B,D)} & -I \end{bmatrix}
\begin{bmatrix}\begin{bmatrix}U_{i,p,\bar{N}}\\Y_{i,p,\bar{N}}\end{bmatrix}\\U_{i_p,f,\bar{N}}\\Y_{i_p,f,\bar{N}}\end{bmatrix}=-H_{(K,I)}E_{{i_p},f,\bar{N}}.\label{eq_Data_equation_1_DeePC_noisee}
\end{eqnarray}
Furthermore, Equation~(\ref{eq_Data_equation_2}) is considered for a noise free prediction:
\begin{eqnarray}
\begin{bmatrix}\Gamma\mathcal{K} & H_{(B,D)} & -I \end{bmatrix}
\begin{bmatrix}\begin{bmatrix}U_{\hat{i},p,1}\\Y_{\hat{i},p,1}\end{bmatrix}\\\color{black}\color{black}U_{\hat{i}_p,f,1}\color{black}\\\color{black}Y_{\hat{i}_p,f,1}\color{black}\color{black}\end{bmatrix}=o.\label{eq_Data_equation_2_DeePC_noisee}
\end{eqnarray}
An instrumental variable $Z_{\bar{N}} \in \mathbb{R}^{q \times \bar{N}} $ is defined such that:
\begin{equation}
   \lim_{\bar{N}\to\infty}\frac{1}{\bar{N}} \left(E_{i_p,f,\bar{N}}Z_{\bar{N}}^T\right)=O,
\end{equation}
and
\begin{equation}
    {\rm{rank}}\left(\lim_{\bar{N}\to\infty}\frac{1}{\bar{N}}\begin{bmatrix}U_{i,p,\bar{N}}\\Y_{i,p,\bar{N}}\\U_{i_p,f,\bar{N}}\end{bmatrix}Z_{\bar{N}}^T\right)={\rm{rank}}\left(\begin{bmatrix}U_{i,p,\bar{N}}\\Y_{i,p,\bar{N}}\\U_{i_p,f,\bar{N}}\end{bmatrix}\right).\label{eq_IV_cond2}
\end{equation}
That is, $Z_{\bar{N}}$ is chosen to be uncorrelated with the block-Hankel matrix containing the noise but highly correlated with the data matrix. A suitable candidate (a variation of the IVs used in PO-MOESP~\cite{verhaegen2007filtering}) is given by:
\begin{equation}
    Z_{\bar{N}}=\begin{bmatrix}U_{i,p,\bar{N}}\\Y_{i,p,\bar{N}}\\U_{i_p,f,\bar{N}}\end{bmatrix},
\end{equation}
which under persistency of excitation conditions~\cite{van2020willems} satisfies the posed conditions~(proof along the lines of IVs for PO-MOESP~\cite{verhaegen2007filtering},~Chapt.~9.6).
%Note that, in the case of closed-loop control the matrices $U_{i_p,f,\bar{N}}$ and $E_{i_p,f,\bar{N}}$ will be correlated and the $U_{i_p,f,\bar{N}}$ in the instrumental variables should be replaced by an external perturbation signal (we will demonstrate this in the simulation section).

Equation~(\ref{eq_Data_equation_1_DeePC_noisee}) is multiplied with the transpose of the instrumental variable and a vector $\color{black}\hat{g}\color{black} \in \mathbb{R}^{q \times 1}$ and (\ref{eq_Data_equation_2_DeePC_noisee}) is subtracted. This leads to the following result:
\begin{eqnarray}
\lim_{\bar{N}\to\infty}\begin{bmatrix}\Gamma\mathcal{K} & H_{(B,D)} & -I \end{bmatrix}\hdots \hspace{3cm} \\
\left(\begin{bmatrix}\begin{bmatrix}U_{i,p,\bar{N}}\\Y_{i,p,\bar{N}}\end{bmatrix}\\U_{i_p,f,\bar{N}}\\Y_{i_p,f,\bar{N}}\end{bmatrix}Z_{\bar{N}}^T\color{black}\hat{g}\color{black}-\begin{bmatrix}\begin{bmatrix}U_{\hat{i},p,1}\\Y_{\hat{i},p,1}\end{bmatrix}\\\color{black}\color{black}U_{\hat{i}_p,f,1}\color{black}\\\color{black}Y_{\hat{i}_p,f,1}\color{black}\color{black}\end{bmatrix}\right)=o,\label{eq_Data_equation_1_DeePC2_noisee}
\nonumber \end{eqnarray}
and similar as in the noise free case:
\begin{eqnarray}
\lim_{\bar{N}\to\infty}\left(\begin{bmatrix}\begin{bmatrix}U_{i,p,\bar{N}}\\Y_{i,p,\bar{N}}\end{bmatrix}\\U_{i_p,f,\bar{N}}\end{bmatrix}Z_{\bar{N}}^T\color{black}\hat{g}\color{black}\right)=\begin{bmatrix}\begin{bmatrix}U_{\hat{i},p,1}\\Y_{\hat{i},p,1}\end{bmatrix}\\\color{black}U_{\hat{i}_p,f,1}\color{black}\end{bmatrix},\label{eq_char_1_DeePC_IV}
\end{eqnarray}
which can be used in a data-enabled prediction control problem with $\color{black}\color{black}\hat{g}\color{black}\color{black}$, $\color{black}U_{\hat{i}_p,f,1}\color{black}$ as the decision variables, and
\begin{eqnarray}
\color{black}Y_{\hat{i}_p,f,1}\color{black}= \lim_{\bar{N}\to\infty}Y_{i_p,f,\bar{N}} Z_{\bar{N}}^T\color{black}\hat{g}\color{black}, \label{eq_char_2_DeePC_IV}
\end{eqnarray}
being an asymptotically noise free prediction of the output. In the nominal DeePC algorithm~\cite{coulson2019data} the predictions are made on linear combinations of noisy output signals.
Adding regularization to the objective function of the nominal DeePC~\cite{coulson2021distributionally} or data-averaging~\cite{jo2022data} can mitigate the effect of the noise.
%\begin{eqnarray}
%\color{black}Y_{\hat{i}_p,f,1}\color{black}=Y_{i_p,f,\bar{N}} %\color{black}g\color{black} %-H_{(K,I)}E_{{i}_{p},f,\bar{N}}\color{black}g\color{black}.
%\end{eqnarray}
%Using sparsity promoting regularisation such as $||g||_0$ or relaxations thereof would emphasise the effect of noise while $||g||_2$ would have an averaging effect on the noise term.
%Also observe that even with noise \eqref{eq_Data_equation_1_DeePC4} is consistent under the condition that $\tilde{A}^p=O$ and there is no need to relax this equation.

%With the chosen dimensional of the instrumental variables the data matrix becomes square with $(r+\ell)p+rf$ columns and rows. Assuming persistency of excitation this matrix is of full rank and an unique relation can be found between the future inputs and outputs.

%The question on how the characteristic equations \eqref{eq_char_1_DeePC_IV}-\eqref{eq_char_2_DeePC_IV} and its solution is different from subspace predictive control~\cite{favoreel1999spc}, is the subject of the next subsection.

\subsection{Subspace predictive control\label{sect:SPC}}
\noindent In this section, the direct equivalence between the Subspace Predictive Control (SPC) and the DeePC algorithm is established through the use of the presented instrumental variables. This further strengthens the results in~\cite{dorfler2021bridging} where an initial connection has been established between SPC and the traditional DeePC algorithm.

Since the data matrix in \eqref{eq_char_1_DeePC_IV} is known and under persistency of excitation conditions invertible, $\color{black}\hat{g}\color{black}$ can be explicitly solved for. Substitution of the obtained expression in~\eqref{eq_char_2_DeePC_IV} results in:
\begin{eqnarray}\color{black}Y_{\hat{i}_p,f,1}\color{black}=Y_{i_p,f,\bar{N}}Z_{\bar{N}}^T(Z_{\bar{N}}Z_{\bar{N}}^T)^{-1}\begin{bmatrix}\begin{bmatrix}U_{\hat{i},p,1}\\Y_{\hat{i},p,1}\end{bmatrix}\\\color{black}U_{\hat{i}_p,f,1}\color{black}\end{bmatrix}.\label{eq_SPC1}
\end{eqnarray}
In the SPC algorithm, a similar solution can be obtained by directly solving~(\ref{eq_Data_equation_1}) for $\begin{bmatrix}\Gamma\mathcal{K} & H_{(B,D)}\end{bmatrix}$ in a least squares sense and putting the available data in a data matrix $Z_{\bar{N}}$:
\begin{equation}
    \min_{\begin{bmatrix}\Gamma\mathcal{K} & H_{(B,D)}\end{bmatrix}} \left|\left| Y_{i_p,f,\bar{N}}-\begin{bmatrix}\Gamma\mathcal{K} & H_{(B,D)}\end{bmatrix}Z_{\bar{N}} \right|\right|_F.
\end{equation}
The solution of this problem is given by the following estimate:
\begin{equation}
    \begin{bmatrix}\hat{\Gamma\mathcal{K}} & \hat{H}_{(B,D)}\end{bmatrix}=Y_{i_p,f,\bar{N}}Z_{\bar{N}}^T(Z_{\bar{N}}Z_{\bar{N}}^T)^{-1}.
\end{equation}
Substitution of this result in (\ref{eq_Data_equation_2}), and taking the expectation of the future noise, directly leads to~(\ref{eq_SPC1}) which shows the direct equivalence.

%The obtained solution, (\ref{eq_SPC1}) , can also be  written in a similar form as in~\cite{favoreel1999spc}:
%\begin{eqnarray}\color{black}Y_{\hat{i}_p,f,1}\color{black}=L_w\begin{bmatrix}U_{\hat{i},p,1}\\Y_{\hat{i},p,1}\end{bmatrix}+L_u\color{black}U_{\hat{i}_p,f,1}\color{black},\label{eq_SPC2}
%\end{eqnarray}
%with $L_w \in \matbb{R}^{\ell f \times (r+\ell)p}$ and $L_u \in %\matbb{R}^{\ell f \times rf}$.

%In~\cite{favoreel1999spc} a QR-factorization is used to obtain the matrices $L_w$ and $L_u$.
The SPC algorithm is also discussed in~\cite{dorfler2021bridging}, where it is also observed that the low-rank property of the estimated $\Gamma \mathcal{K}$ matrix, and the lower-triangular block-Toeplitz property of the estimated $H_{(B,D)}$ matrix are not enforced.  In~\cite{dorfler2021bridging}, methods are proposed to enforce this structure by the employment of regularization techniques.

%In~\cite{dong2008closed} the idea of Favoreel is extended by exploiting the structure in the data equations to enforce the causality condition and solve the closed-loop issue. This resulted in a more efficient use of data and thus faster convergence. In the field of closed-loop SPC also extensions exist for Hammerstein~\cite{kulcsar2009closed} and LPV~\cite{dong2009closed} model structures and for repetitive control~\cite{navalkar2014subspace} or cautious data-driven control~\cite{dong2009cautious}.

\begin{figure*}[htp!]
    \centering
    \includegraphics[width = 1.00\textwidth]{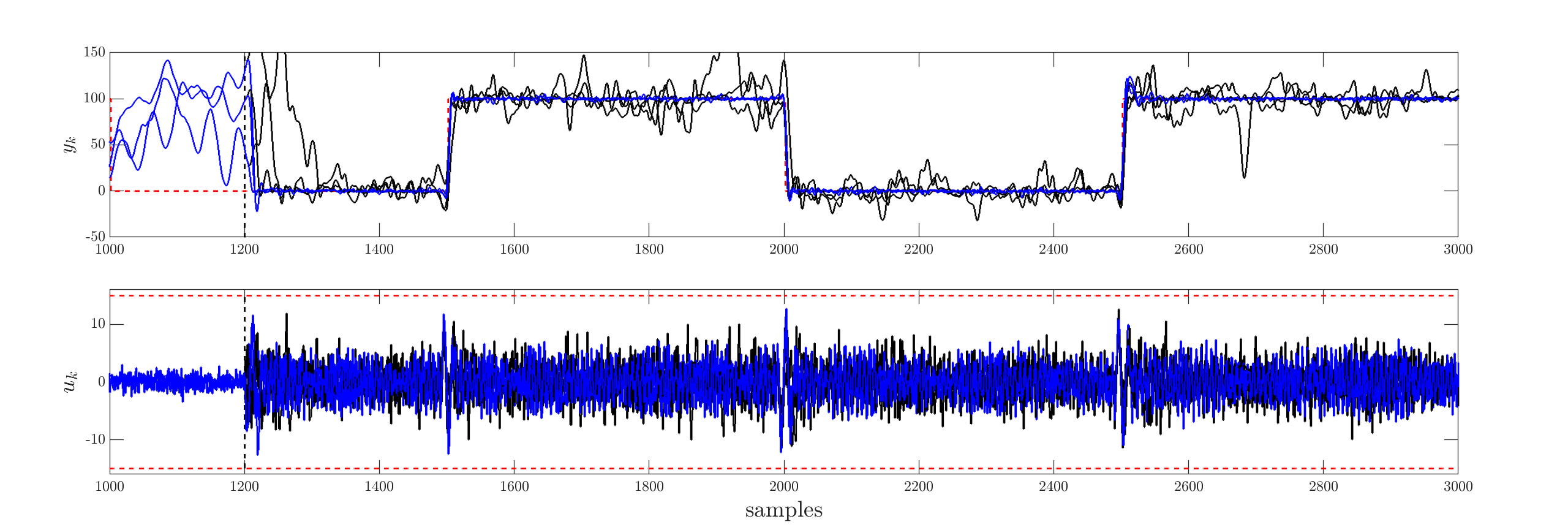}
    \caption{Simulation results comparing the reference tracking performance of DeePC with averaging (black) and DeePC with IV (blue). For~$p=f=20$ with~$\rm{var}(e_k)=0.1^2$ and~$\bar{N}=500$, the proposed IV method outperforms the original method in terms of the tracking error for three distinct realizations.}
        \label{fig:effect_of_N}
                \vspace*{0.0cm}
\end{figure*}

%%%%%%%%%%%%%%%%%%%%%%%%%%%%%%%%%%%%%%%%%%%%%%%%%%%%%%%%%%%%%%%%%%%%%%%%
\section{DeePC \label{Section_Predictive_Control}}
\noindent With the equations from the previous section at hand, the original DeePC problem as proposed in \cite{coulson2019data} is defined as:
\begin{align}
\min_{\color{black}U_{\hat{i}_p,f,1}\color{black},\color{black}g\color{black}}              && \color{black}Y_{\hat{i}_p,f,1}\color{black}^TQ\color{black}Y_{\hat{i}_p,f,1}\color{black}&\,+\,\color{black}U_{\hat{i}_p,f,1}\color{black}^TR\color{black}U_{\hat{i}_p,f,1}\color{black}  \nonumber\\
\text{subject to: } &&  \begin{bmatrix}U_{i,p,\bar{N}}\\Y_{i,p,\bar{N}}\\U_{i_p,f,\bar{N}}\end{bmatrix}\color{black}g\color{black}&=\begin{bmatrix}U_{\hat{i},p,1}\\Y_{\hat{i},p,1}\\\color{black}U_{\hat{i}_p,f,1}\color{black}\\\end{bmatrix},\\
&& \color{black}Y_{\hat{i}_p,f,1}\color{black}&=Y_{i_p,f,\bar{N}} \color{black} g  \color{black}\nonumber.
\end{align}
The cost-function can be extended with a regularization term (\emph{e.g.}, $\lambda ||g||_2$), or input/output constraints. With the addition of IVs to the DeePC problem, a slightly modified optimization problem is defined:
\begin{align}
\min_{\color{black}U_{\hat{i}_p,f,1}\color{black},\color{black}\hat{g}\color{black}}              && \color{black}Y_{\hat{i}_p,f,1}\color{black}^TQ\color{black}Y_{\hat{i}_p,f,1}\color{black}&\,+\,\color{black}U_{\hat{i}_p,f,1}\color{black}^TR\color{black}U_{\hat{i}_p,f,1}\color{black}  \nonumber\\
\text{subject to: } &&  \begin{bmatrix}U_{i,p,\bar{N}}\\Y_{i,p,\bar{N}}\\U_{i_p,f,\bar{N}}\end{bmatrix}Z_{\bar{N}}^T\color{black}\hat{g}\color{black}&=\begin{bmatrix}U_{\hat{i},p,1}\\Y_{\hat{i},p,1}\\\color{black}U_{\hat{i}_p,f,1}\color{black}\end{bmatrix}\\
&& Y_{i_p,f,\bar{N}}Z_{\bar{N}}^T\color{black}\hat{g}\color{black} &=\color{black}Y_{\hat{i}_p,f,1}\color{black} \nonumber
\end{align}
where $\color{black}\hat{g}\color{black}$ can be seen as a dummy variable and explicitly solved for as explained in the previous section.
Note that in the noiseless case, the matrix $Z_{\bar{N}}Z_{\bar{N}}^T$ becomes singular and regularization has to be applied or a different data window has to be chosen (different $f$ and/or $p$).

%%%%%%%%%%%%%%%%%%%%%%%%%%%%%%%%%%%%%%%%%%%%%%%%%%%%%%%%%%%%%%%%%%%%%%%%

\section{Simulation study \label{Sec_Simulations}}
\noindent In this section, the presented IV-based DeePC algorithm is compared with the traditional DeePC algorithm for various levels of noise. The simulation study considers a $5^\mathrm{th}$ order system described in~\cite{favoreel1999spc}, representing an actuated laboratory test setup with two circular plates and flexible shafts. The discrete-time system is compatible with the model structure in~(\ref{eqmod}), and the corresponding system matrices are given:
\begin{alignat*}{2}
&A=\begin{bmatrix}
4.4 & 1 & 0 & 0 & 0\\ -8.09 &0 & 1 & 0& 0\\ 7.83 & 0 & 0 &1 &0\\ -4 &0 &0 & 0 & 1\\ 0.86 & 0 &0 &0 &0
\end{bmatrix}, \hspace{2mm} &&B=\begin{bmatrix}0.00098\\ 0.01299\\ 0.01859\\ 0.0033\\ -0.00002\end{bmatrix}\\
&C=\begin{bmatrix}
1& 0& 0& 0& 0
\end{bmatrix}, \hspace{2mm} &&K=\begin{bmatrix}2.3\\ -6.64\\ 7.515\\ -4.0146\\ 0.86336\end{bmatrix}, \hspace{2mm} D=0.
\end{alignat*}
The objective is to track a square wave with an amplitude of 50 and a DC-offset of 50. During the first $1200$ samples the system is excited with a normally distributed white noise signal with a variance of $1$. After this period, the excitation is removed and the data-enabled predictive controllers are enabled in a receding horizon setting. The input magnitude and rate of variation are respectively constrained to ${|u_k|\leq 15}$ and ${|u_{k+1}-u_k|\leq 3.75}$.

Fig.~\ref{fig:effect_of_N} shows simulation results of the output and input trajectories of DeePC with IV and random averaging. In the case of random averaging, the matrix $Z_{\bar{N}} \in \mathbb{R}^{\left(r+\ell\right)p+rf \times \bar{N} }$ is chosen randomly. The prediction window is chosen as ${f=20}$ and $p$ is selected to equal $f$ (common choice in subspace identification~\cite{van2013closed}), the noise variance is taken as~${\rm{var}(e_k)}=0.1^2$, and the data window~$\bar{N}=500$. It is observed that the tracking performance of the proposed method outperforms the original method for three different realisations.

The simulation studies in the subsequent subsections consider the effect of the data window~$\left(\bar{N}\right)$, noise level~$\left(\rm{var}{(e_k)}\right)$ and the prediction window~$\left(f\right)$, and are presented in Fig.~\ref{fig:simulation_cases}. The performance is measured in terms of the mean squared error of the tracking error divided by the two norm of the reference signal.

\subsection{The data window length $\bar{N}$}
\noindent The tracking error performance of both DeePC approaches for different sizes of $\bar{N}$ is shown in Fig.~\ref{fig:compare_IV_vs_random_averaging}. For each $\bar{N}$, $100$ distinct noise realisations are taken.
Since $p=f=20$, the data matrix is square for $\bar{N}=60$. In the case of $\bar{N}\leq 60$, there is no clear difference between the proposed and the original method. However, increased data windows ($\bar{N}>60$) clearly demonstrate the strength of the proposed IV approach with increased tracking performance, whereas the performance of the original algorithm deteriorates for larger values of $\bar{N}$.
% \begin{figure}[b!]
%     \centering
%     \includegraphics[width = 1.00\columnwidth]{Figures/Results_Np_noisePE_0_0_input_pinv_e_0_5_Np_30_2_300_f_20_p_20.png}
%     \caption{The tracking error for the DeePC with instrumental variables (blue), and the original DeePC method with random averaging (black) as a function of the data window. For each $\bar{N}$ the simulation is repeated 100 times. The results are obtained for $p=f=20$ with ${\rm{var}}(e_k)=0.5^2$. The solid lines represent the median of the tracking error while the shaded area covers 80\% of the realisations. The noise mitigation properties of the IV method outperform that of random averaging, and the performance increases with increasing $\bar{N}$.}
%     \label{fig:compare_IV_vs_random_averaging}
% \end{figure}

\subsection{The effect of ${\rm{var}}(e_k)$}
\noindent Fig.~\ref{fig:effect_of_e} compares the tracking performance of the two DeePC algorithms for different noise levels. For each $\rm{var}(e_k)$, $100$ distinct noise realisations are taken. The results clearly demonstrate the beneficial effect of the proposed method by outperforming the original algorithm for higher noise levels. For lower variances of the noise, the performance of both approaches is comparable.
% \begin{figure}[h]
%     \centering
%     \includegraphics[width = 1.00\columnwidth]{Figures/Results_effectnoise_Np_noisePE_0_0_input_pinv_e_v_Np_200_f_20_p_20}
%     \caption{The tracking error for the DeePC with instrumental variables (blue), and the original DeePC method with random averaging (black) as a function of the noise level. The simulation is repeated 100 times for different noise realisations, and $p=f=20$ with $\bar{N}=200$. The solid lines represent the median of the tracking error while the shaded area covers $80\%$ of the realisations.
%     \label{fig:effect_of_e}}
% \end{figure}

\subsection{The effect of the prediction window $f$}
\begin{figure}
     \centering
     \begin{subfigure}[b]{1.0\columnwidth}
         \centering
             \includegraphics[width=1.0\columnwidth]{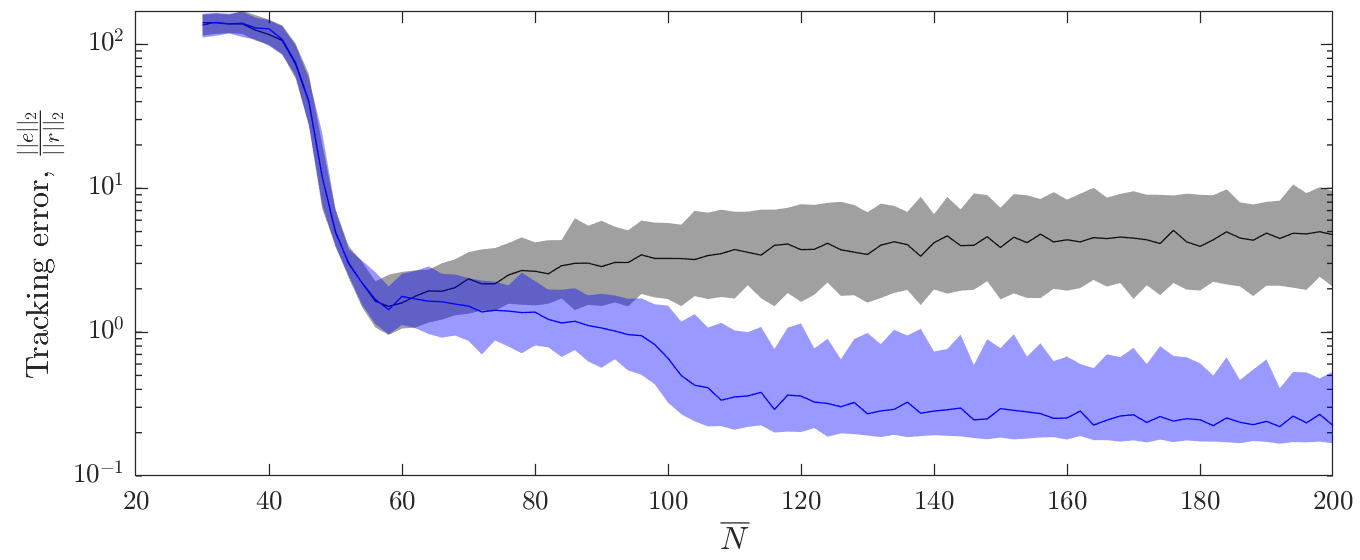}
         \caption{Varying data window $\bar{N}$, and $p=f=20$ with ${\rm{var}}(e_k)=0.5^2$}
         \label{fig:compare_IV_vs_random_averaging}
     \end{subfigure}
     \par\vspace{0.7cm} % force a bit of vertical whitespace
     \begin{subfigure}[b]{1.0\columnwidth}
         \centering
         \includegraphics[width=\columnwidth]{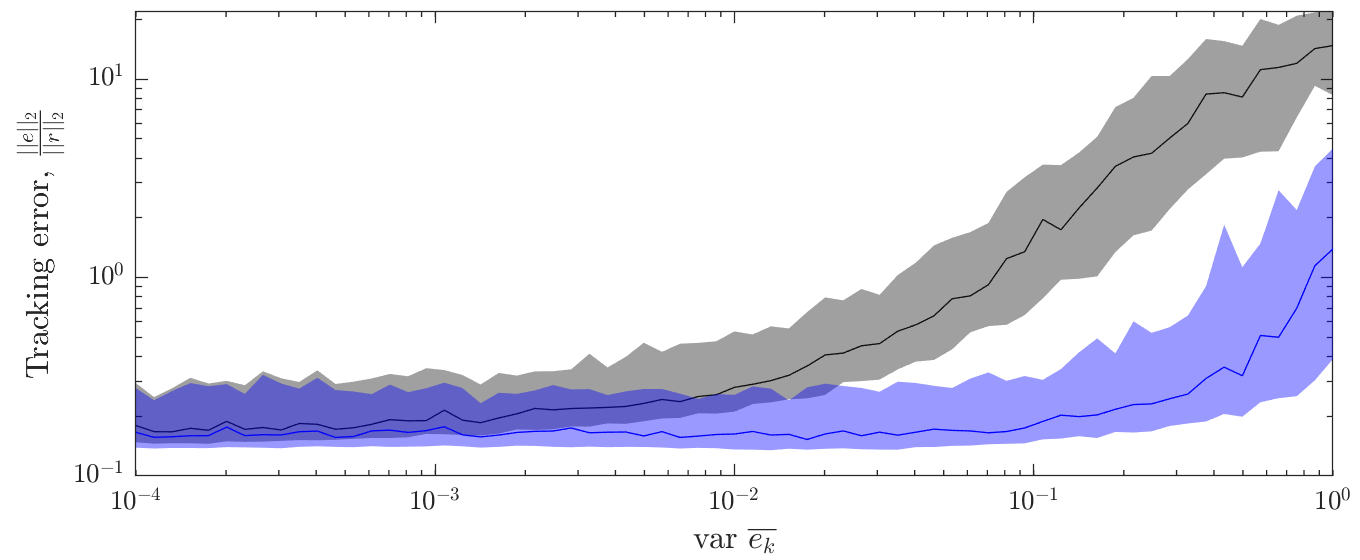}
         \caption{Varying noise level ${\rm{var}}(e_k)$, and $p=f=20$ with $\bar{N}=200$.}
         \label{fig:effect_of_e}
     \end{subfigure}
     \par\vspace{0.7cm} % force a bit of vertical whitespace
     \begin{subfigure}[b]{1.0\columnwidth}
         \centering
              \includegraphics[width=1.0\columnwidth]{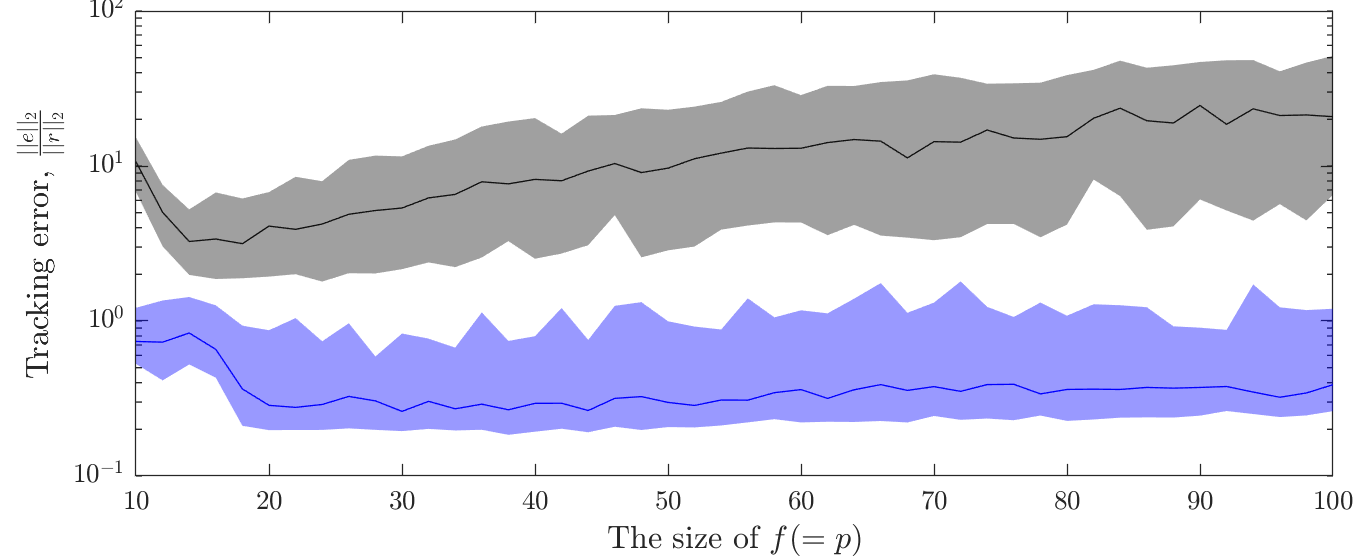}
         \caption{Varying prediction window $f(=p)$, and ${\rm{var}}(e_k)=0.5^2$ with $\bar{N}=6\times f$. }
         \label{fig:effect_of_f}
     \end{subfigure}
        \caption{Performance in terms of tracking error for the DeePC with instrumental variables (blue), and the original DeePC method with random averaging (black). Simulations are repeated 100 times for different noise realisations. The solid lines represent the median of the tracking error while the shaded area covers 80\% of the realisations.}
        \label{fig:simulation_cases}
\end{figure}

\noindent The effect of the prediction window $f$ on the tracking performance for the proposed and original method are presented in Fig.~\ref{fig:effect_of_f}. To mitigate the effect of $\bar{N}$ on the presented results, the data window is chosen as $\bar{N}=6\times f$, and for simplicity $p=f$. For each $f$, $100$ new noise realisations are taken.
% \begin{figure}[b!]
%     \centering
%     \includegraphics[width = 1.00\columnwidth]{Figures/Results_effect_p_noisePE_0_0_input_pinv_e_0_5_f_10_2_100}
%     \caption{The tracking error for the DeePC with instrumental variables (blue), and the original DeePC method with random averaging (black) as a function of $f(=p)$. The simulation is repeated 100 times for different noise realisations, and ${\rm{var}}(e_k)=0.5^2$ with $\bar{N}=6\times f$. The solid lines represent the median of the tracking error while the shaded area covers 80\% of the realisations.
%     \label{fig:effect_of_f}}
% \end{figure}

The results demonstrate that there is a minor overall effect of the prediction window on the performance. It is concluded that the gained performance of the IV method with respect to random averaging is independent from the prediction window~$f$.

\section{CONCLUSIONS \label{Sect_Conclusions}}
\noindent In this paper, a new instrumental variable approach to DeePC-based data-driven control is presented, that leads to a direct equivalence between two existing classes of direct data-driven control methods: Data-enabled Predictive Control (DeePC) and Subspace Predictive Control (SPC). To end up at this result, first the characteristic equation used in DeePC is derived from the data-equation typically used in subspace identification algorithms. Instrumental variables~(IVs) are included in the DeePC framework, with the intent to asymptotically remove the effect of process and measurement noise. A particular choice of IVs is made that is uncorrelated with future noise, but at the same time highly correlated with the data matrix. Simulation studies showcase the improved performance in terms of the tracking error of the proposed algorithm in the case of process and measurement noise.

\bibliographystyle{IEEEtran}
\bibliography{root.bib}

\end{document}